\documentclass[12pt]{article}
  \usepackage{amsfonts}
  \usepackage{amsmath}
\usepackage{amssymb}
\usepackage{amscd}
\usepackage{graphicx}
\begin{document}

~~
\bigskip
\bigskip
\begin{center}
{\Large {\bf{{{Noncommutative Sprott systems and their jerk dynamics}}}}}
\end{center}
\bigskip
\bigskip
\bigskip
\begin{center}
{{\large ${\rm {Marcin\;Daszkiewicz}}$}}
\end{center}
\bigskip
\begin{center}
\bigskip

{ ${\rm{Institute\; of\; Theoretical\; Physics}}$}

{ ${\rm{ University\; of\; Wroclaw\; pl.\; Maxa\; Borna\; 9,\;
50-206\; Wroclaw,\; Poland}}$}

{ ${\rm{ e-mail:\; marcin@ift.uni.wroc.pl}}$}

\end{center}
\bigskip
\bigskip
\bigskip
\bigskip
\bigskip
\bigskip
\bigskip
\bigskip
\bigskip
\begin{abstract}
In this article we provide the noncommutative Sprott models. We demonstrate, that effectively, each of them is described
by system of three complex, ordinary and nonlinear differential equations. Apart of that, we find for such modified models the
corresponding (noncommutative) jerk dynamics as well as we study numerically as an example, the deformed Sprott-A system.
\end{abstract}
\bigskip
\bigskip
\bigskip
\bigskip
\eject

In the last four decades there appeared a lot of papers dealing with so-called chaotic
models, i.e., with such models whose dynamics is described by strongly sensitive with
respect initial conditions, nonlinear differential equations. The most popular of them
are: Lorenz system \cite{[1]}, Roessler system \cite{[2]}, Rayleigh-Benard system \cite{[3]}, Henon-Heiles
system \cite{[4]}, jerk equation \cite{[5]}, Duffing equation \cite{[6]}, Lotka-Volter system \cite{[7]}, Liu system
\cite{[8]}, Chen system \cite{[9]} and Sprott systems \cite{[10]}.

The quite interesting among the above models seem to be
the last ones, i.e., the Sprott systems. They have been provided in article \cite{[10]} by
systematic examination of most simple, three-dimensional and ordinary differential equations, with
quadratic nonlinearities. There has been uncovered 19 examples
of chaotic flows listed in the following {Table}:
\\
\\
\\
\begin{tabular}{ccccc}
\hline $\;\;\;\;{\rm type}\;\;\;\;$ & $\;\;\;\;\;\;\;\;{\rm 1st\;equation}\;\;\;\;\;\;\;$ & $\;\;\;\;\;\;\;\;{\rm 2nd\;equation}\;\;\;\;\;\;\;$ & $\;\;\;\;\;\;\;\;{\rm 3rd\;equation}\;\;\;\;\;\;\;$\\
\hline ${\rm A}$ & $\dot{x}_1=x_2$ & $\dot{x}_2=-x_1+x_2x_3$ & $\dot{x}_3=1-x_2^2$\\
       ${\rm B}$ & $\dot{x}_1=x_2x_3$ & $\dot{x}_2=x_1-x_2$ & $\dot{x}_3=1-x_1x_2$\\
       ${\rm C}$ & $\dot{x}_1=x_2x_3$ & $\dot{x}_2=x_1-x_2$ & $\dot{x}_3=1-x_1^2$\\
       ${\rm D}$ & $\dot{x}_1=-x_2$ & $\dot{x}_2=x_1+x_3$ & $\dot{x}_3=x_1x_3+ 3x_2^2$\\
       ${\rm E}$ & $\dot{x}_1=x_2x_3$ & $\dot{x}_2=x_1^2-x_2$ & $\dot{x}_3=1-4x_1$\\
       ${\rm F}$ & $\dot{x}_1=x_2+x_3$ & $\dot{x}_2=-x_1+0.5x_2$ & $\dot{x}_3=x_1^2-x_3$\\
       ${\rm G}$ & $\dot{x}_1=0.4x_1+x_3$ & $\dot{x}_2=x_1x_3-x_2$ & $\dot{x}_3=-x_1+x_2$\\
       ${\rm H}$ & $\dot{x}_1=-x_2+x_3^2$ & $\dot{x}_2=x_1+0.5x_2$ & $\dot{x}_3=x_1-x_3$\\
       ${\rm I}$ & $\dot{x}_1=-0.2x_2$ & $\dot{x}_2=x_1+x_3$ & $\dot{x}_3=x_1+ x_2^2-x_3$\\
       ${\rm J}$ & $\dot{x}_1=2x_3$ & $\dot{x}_2=-2x_2+x_3$ & $\dot{x}_3=-x_1+x_2+x_2^2$\\
       ${\rm K}$ & $\dot{x}_1=x_1x_2 -x_3$ & $\dot{x}_2=x_1-x_2$ & $\dot{x}_3=x_1+0.3x_3$\\
       ${\rm L}$ & $\dot{x}_1=x_2+3.9x_3$ & $\dot{x}_2=0.9x_1^2-x_2$ & $\dot{x}_3=1-x_1$\\
       ${\rm M}$ & $\dot{x}_1=-x_3$ & $\dot{x}_2=-x_1^2-x_2$ & $\dot{x}_3=1.7+1.7x_1+x_2$\\
       ${\rm N}$ & $\dot{x}_1=-2x_2$ & $\dot{x}_2=x_1+x_3^2$ & $\dot{x}_3=1+x_2-2x_1$\\
       ${\rm O}$ & $\dot{x}_1=x_2$ & $\dot{x}_2=x_1-x_3$ & $\dot{x}_3=x_1+x_1x_3+2.7x_2$\\
       ${\rm P}$ & $\dot{x}_1=2.7x_2+x_3$ & $\dot{x}_2=-x_1+x_2^2$ & $\dot{x}_3=x_1+x_2$\\
       ${\rm Q}$ & $\dot{x}_1=-x_3$ & $\dot{x}_2=x_1-x_2$ & $\dot{x}_3=3.1x_1+x_2^2+0.5x_3$\\
       ${\rm R}$ & $\dot{x}_1=0.9-x_2$ & $\dot{x}_2=0.4+x_3$ & $\dot{x}_3=x_1x_2-x_3$\\
       ${\rm S}$ & $\dot{x}_1=-x_1-4x_2$ & $\dot{x}_2=x_1+x_3^2$ & $\dot{x}_3=1+x_1$\\
\hline
\end{tabular}

\begin{center}
{\bf Table 1.} The Sprott systems.
\end{center}
~~\\
The above models have been applied, for example, in Electronics \cite{a1}, Physics \cite{a2} and Chemistry \cite{a3}. Besides,
the corresponding jerk dynamics were introduced in paper \cite{jerksprott} by using of the proper nonlinear transformation rules of variables
$x_1$, $x_2$ and $x_3$; they look as follows:
\\
\\
\\
\begin{tabular}{ccccc}
\hline $\;\;{\rm type}\;\;$ & $\;\;\;\;\;\;\;\;\;\;\;\;\;\;\;\;\;\;\;\;\;\;\;\;\;\;\;\;\;\;\;\;\;\;\;\;\;\;\;\;\;\;\;\;\;
{\rm jerk\;dynamics}\;\;\;\;\;\;\;\;\;\;\;\;\;\;\;\;\;\;\;\;\;\;\;\;\;\;\;\;\;\;\;\;\;\;\;\;\;\;\;\;\;\;\;\;\;$ &  $$\\
\hline ${\rm A}$ & $-----------$ \\
       ${\rm B}$ & $-----------$ \\
       ${\rm C}$ & $-----------$ \\
       ${\rm D}$ & $\dddot{x}_1=x_1\ddot{x}_1-\dot{x}_1-3\dot{x}_1^2+x_1^2$ \\
       ${\rm E}$ & $-----------$ \\
       ${\rm F}$ & $\dddot{x}_1=-0.5\ddot{x}_1-0.5\dot{x}_1-0.5x^2_1+2\dot{x}_1x_1-x_1$\\
       ${\rm G}$ & $\dddot{x}_1=-0.6\ddot{x}_1-0.6\dot{x}_1-0.4x^2_1+\dot{x}_1x_1-x_1$\\
       ${\rm H}$ & $\dddot{x}_3=-0.5\ddot{x}_3-0.5\dot{x}_3-0.5x^2_3+2\dot{x}_3x_3-x_3$\\
       ${\rm I}$ & $\dddot{x}_1=-\ddot{x}_1-0.2\dot{x}_1-5\dot{x}_1^2-0.4x_1$\\
       ${\rm J}$ & $\dddot{x}_2=-2\ddot{x}_2-\dot{x}_2-4x_2+2\dot{x}_2x_2$\\
       ${\rm K}$ & $\dddot{x}_2=-0.7\ddot{x}_2-0.7\dot{x}_2-x_2+x_2\ddot{x}_2+1.7x_2\dot{x}_2-0.3x_2^2+\dot{x}_2^2$\\
       ${\rm L}$ & $\dddot{x}_1=-\ddot{x}_1-3.9\dot{x}_1-3.9x_1+1.8x_1\dot{x}_1+3.9$\\
       ${\rm M}$ & $\dddot{x}_1=-\ddot{x}_1-1.7\dot{x}_1-1.7x_1+x_1^2-1.7$\\
       ${\rm N}$ & $\dddot{x}_3=-2\ddot{x}_3-2\dot{x}_3-4x_3+2\dot{x}_3x_3+2$\\
       ${\rm O}$ & $\dddot{x}_1= x_1\ddot{x}_1-1.7\dot{x}_1-x_1-x_1^2$\\
       ${\rm P}$ & $\dddot{x}_2=2x_2\ddot{x}_2-1.7\dot{x}_2+2\dot{x}_2^2-x_2-x_2^2$\\
       ${\rm Q}$ & $\dddot{x}_2=-0.5\ddot{x}_2-2.6\dot{x}_2-3.1x_2-x_2^2$\\
       ${\rm R}$ & $\dddot{x}_1=-\ddot{x}_1-0.9x_1 +x_1\dot{x}_1-0.4$\\
       ${\rm S}$ & $\dddot{x}_3=-\ddot{x}_3-4\dot{x}_3-4x_3^2+4$\\
\hline
\end{tabular}

\begin{center}
{\bf Table 2a.} The jerk dynamics.
\end{center}
\begin{tabular}{ccccc}
\hline $\;\;{\rm type}\;\;$ & $\;\;\;\;\;\;\;\;\;\;\;\;\;\;\;\;\;\;\;\;\;\;\;\;\;\;\;\;\;\;\;\;\;\;\;\;\;\;\;
{\rm remaining\;trajectories}\;\;\;\;\;\;\;\;\;\;\;\;\;\;\;\;\;\;\;\;\;\;\;\;\;\;\;\;\;\;\;\;\;\;\;\;\;\;\;$ &  $$\\
\hline ${\rm A}$ & $-----------$ \\
       ${\rm B}$ & $-----------$ \\
       ${\rm C}$ & $-----------$ \\
       ${\rm D}$ & $x_2=-\dot{x}_1\;\;\;\;\;\;\;\;\;\;\;\;\;\;\;\;\;\;\;\;\;\;\;\;\;\;\;\;\;\;\;\;\;x_3=-\ddot{x}_1-x_1$ \\
       ${\rm E}$ & $-----------$ \\
       ${\rm F}$ & ${x}_2=0.6(\ddot{x}_1+\dot{x}_1+x_1-{x}_1^2)\;\;\;\;\;\;\;\;\;\;{x}_3=0.6(-\ddot{x}_1+0.5\dot{x}_1-x_1+{x}_1^2)$\\
       ${\rm G}$ & ${x}_2=\ddot{x}_1-0.4\dot{x}_1+x_1\;\;\;\;\;\;\;\;\;\;\;\;\;\;\;\;\;\;\;\;\;\;\;\;\;\;x_3=\dot{x}_1-0.4x_1\;\;\;\;\;\;\;\;\;$\\
       ${\rm H}$ & ${x}_1=\dot{x}_3+{x}_3\;\;\;\;\;\;\;\;\;\;\;\;\;\;\;\;\;\;\;\;\;\;\;\;\;\;\;\;\;x_2=-\ddot{x}_3-\dot{x}_3+x^2_3$\\
       ${\rm I}$ & $\;\;\;\;\;{x}_2=-5\dot{x}_1\;\;\;\;\;\;\;\;\;\;\;\;\;\;\;\;\;\;\;\;\;\;\;\;\;\;\;\;\;\;\;\;\;\;\;\;x_3=-5\ddot{x}_1-x_1$\;\;\;\;\;\;\;\;\;\\
       ${\rm J}$ & $\;\;\;{x}_1=-\ddot{x}_2-2\dot{x}_2+x_2+{x}_2^2\;\;\;\;\;\;\;\;\;\;\;\;\;\;\;\;\;\;\;\;\;\;\;\;\;\;\;x_3=2x_2+\dot{x}_2\;\;\;\;\;\;\;\;\;\;\;\;\;\;\;\;$\\
       ${\rm K}$ & $\;\;\;\;\;{x}_1=\dot{x}_2+{x}_2\;\;\;\;\;\;\;\;\;\;\;\;\;\;\;\;\;\;\;\;\;\;\;\;\;\;\;\;\;\;\;x_3=-\ddot{x}_2-\dot{x}_2+x_2\dot{x}_2+x_2^2$\\
       ${\rm L}$ & ${x}_2=-\ddot{x}_1-3.9{x}_1+0.9x_1^2+3.9\;\;\;\;\;\;\;\;\;\;\;\;\;\;\;x_3=0.2(\ddot{x}_1+\dot{x}_1+3.9x_1\;+$\\
       $~~$ & $\;\;\;\;\;\;\;\;\;\;\;\;\;\;\;\;\;\;\;\;\;\;\;\;\;\;\;\;\;\;\;\;\;\;\;\;\;\;\;\;\;\;\;\;\;\;\;\;\;\;\;\;\;\;\;\;\;\;-\;0.9x_1^2-3.9)$\\
       ${\rm M}$ & ${x}_2=-\ddot{x}_1-1.7x_1-1.7\;\;\;\;\;\;\;\;\;\;\;\;\;\;\;\;\;\;\;\;\;\;\;\;x_3=-\dot{x}_1\;\;\;\;\;\;\;\;\;\;\;\;\;\;$\\
       ${\rm N}$ & ${x}_1=\ddot{x}_3+2\dot{x}_3-{x}_3^2\;\;\;\;\;\;\;\;\;\;\;\;\;\;\;\;\;\;\;\;\;\;\;\;\;\;\;\;\;\;x_2=\dot{x}_3+2x_3-1$\\
       ${\rm O}$ & $\;\;\;\;\;{x}_2= \dot{x}_1\;\;\;\;\;\;\;\;\;\;\;\;\;\;\;\;\;\;\;\;\;\;\;\;\;\;\;\;\;\;\;\;\;\;\;\;x_3=-\ddot{x}_1+x_1$\;\;\;\;\;\;\;\\
       ${\rm P}$ & $\;\;\;\;\;\;{x}_1=-\dot{x}_2+x_2^2\;\;\;\;\;\;\;\;\;\;\;\;\;\;\;\;\;\;\;\;\;\;\;\;\;\;\;\;\;\;\;\;\;x_3=-\ddot{x}_2+2x_2\dot{x}_2-2.7x_2$\\
       ${\rm Q}$ & $\;\;\;\;\;{x}_1=\dot{x}_2+{x}_2\;\;\;\;\;\;\;\;\;\;\;\;\;\;\;\;\;\;\;\;\;\;\;\;\;\;\;\;\;\;\;\;\;\;\;\;x_3=-\ddot{x}_2-\dot{x}_2$\;\;\;\;\;\;\;\;\;\\
       ${\rm R}$ & $\;\;\;\;\;{x}_2=-\dot{x}_1+0.9\;\;\;\;\;\;\;\;\;\;\;\;\;\;\;\;\;\;\;\;\;\;\;\;\;\;\;\;\;\;\;\;\;\;\;\;x_3=-\ddot{x}_1-0.4$\;\;\;\;\;\;\;\;\;\\
       ${\rm S}$ & $\;\;\;\;\;\;\;\;\;\;{x}_1=\dot{x}_3-1\;\;\;\;\;\;\;\;\;\;\;\;\;\;\;\;\;\;\;\;\;\;\;\;\;\;\;\;\;\;\;\;\;x_2=0.25(-\ddot{x}_3-\dot{x}_3+1)$\\
\hline
\end{tabular}

\begin{center}
{\bf Table 2b.} The jerk dynamics - remaining trajectories.
\end{center}

In this article we define the Sprott systems on noncommutative space of the form
\begin{eqnarray}
\left[\;\hat{x}_i,\hat{x}_j\;\right] = if_{ij}(t) = if(t)\epsilon_{ij} \;, \label{eq1}
\end{eqnarray}
with arbitrary time-dependent real functions $f(t)$ and antisymmetric tensor $\epsilon_{ij}=-\epsilon_{ji}$\footnote{One should notice, that the commutation relations (\ref{eq1}) reproduce for constant function $f_{ij}(t) = \theta_{ij}$ the well-known, most simple version of deformed space, called canonical type of space-time noncommutativity \cite{oeckl}-\cite{chai}.}. The motivations for such studies are quite obvious and natural, i.e., it gives the general information on the impact of high-energy (just noncommutative) regime on the one of the most popular chaotic (nonlinear) models. It should be noted, however, that such kind of investigations has been already performed for example in articles \cite{lit1} and \cite{lit2}.

In the first step of our construction we remind, that the commutation relationes (\ref{eq1}) can be (formally) realized in the framework of Quantum Group Theory \cite{qg}, with use so-called twist procedure \cite{twist}. Then, the quantum space is represented by Hopf module  equipped with the following $\star$-product for two arbitrary classical functions $f(x)$ and $g(x)$ \cite{hopf}
\begin{eqnarray}
f(x)\star g(y) = f(x)\exp\left(\frac{i}{2}f_{ij}(t)\partial^i_x\otimes \partial^j_y\right)g(y)\;. \label{eq2}
\end{eqnarray}
Particularly, for $f(x)=x$ and $g(y)=y$ we have
\begin{eqnarray}
x_i\star x_j &=& x_i\exp\left(\frac{i}{2}f_{kl}(t)\partial^k_x\otimes \partial^l_x\right)x_j = x_ix_j + i\frac{f_{ij}(t)}{2}\;, \label{eq3}\\
x_j\star x_i &=& x_j\exp\left(\frac{i}{2}f_{kl}(t)\partial^k_x\otimes \partial^l_x\right)x_i = x_jx_i + i\frac{f_{ji}(t)}{2}\;, \label{eq4}
\end{eqnarray}
and, consequently
\begin{eqnarray}
\left[\;\hat{x}_i,\hat{x}_j\;\right] &=& \left[\;{x}_i,{x}_j\;\right]_\star =  x_i\star x_j - x_j\star x_i = i{f_{ij}}(t)\;. \label{eq5}
\end{eqnarray}
It should be noted, that quantum space (\ref{eq1}) has been explicitly constructed only for particular form of function $f_{ij}(t)$ in papers \cite{hopf}, \cite{kon2} and \cite{kon3}.

Next, we define the noncommutative Sprott systems by the following replacement
\begin{eqnarray}
x_i x_j \to x_i\star x_j = x_ix_j + i\frac{f_{ij}(t)}{2}\;\;;\;\;i \ne j\;, \label{eq6}
\end{eqnarray}
provided in all dynamical equations from {\bf Table 1}. In such a way, we get:
\\
\\
\\
\begin{tabular}{ccccc}
\hline $\;\;\;\;{\rm type}\;\;\;\;$ & $\;\;\;\;\;\;\;\;{\rm 1st\;equation}\;\;\;\;\;\;\;$ & $\;\;\;\;\;\;\;\;{\rm 2nd\;equation}\;\;\;\;\;\;\;$ & $\;\;\;\;\;\;\;\;{\rm 3rd\;equation}\;\;\;\;\;\;\;$\\
\hline ${\rm A}$ & $\dot{x}_1=x_2$ & $\dot{x}_2=-x_1+x_2\star x_3$ & $\dot{x}_3=1-x_2^2$\\
       ${\rm B}$ & $\dot{x}_1=x_2\star x_3$ & $\dot{x}_2=x_1-x_2$ & $\dot{x}_3=1-x_1\star x_2$\\
       ${\rm C}$ & $\dot{x}_1=x_2\star x_3$ & $\dot{x}_2=x_1-x_2$ & $\dot{x}_3=1-x_1^2$\\
       ${\rm D}$ & $\dot{x}_1=-x_2$ & $\dot{x}_2=x_1+x_3$ & $\dot{x}_3=x_1\star x_3+ 3x_2^2$\\
       ${\rm E}$ & $\dot{x}_1=x_2\star x_3$ & $\dot{x}_2=x_1^2-x_2$ & $\dot{x}_3=1-4x_1$\\
       ${\rm F}$ & $\dot{x}_1=x_2+x_3$ & $\dot{x}_2=-x_1+0.5x_2$ & $\dot{x}_3=x_1^2-x_3$\\
       ${\rm G}$ & $\dot{x}_1=0.4x_1+x_3$ & $\dot{x}_2=x_1\star x_3-x_2$ & $\dot{x}_3=-x_1+x_2$\\
       ${\rm H}$ & $\dot{x}_1=-x_2+x_3^2$ & $\dot{x}_2=x_1+0.5x_2$ & $\dot{x}_3=x_1-x_3$\\
       ${\rm I}$ & $\dot{x}_1=-0.2x_2$ & $\dot{x}_2=x_1+x_3$ & $\dot{x}_3=x_1+ x_2^2-x_3$\\
       ${\rm J}$ & $\dot{x}_1=2x_3$ & $\dot{x}_2=-2x_2+x_3$ & $\dot{x}_3=-x_1+x_2+x_2^2$\\
       ${\rm K}$ & $\dot{x}_1=x_1\star x_2 -x_3$ & $\dot{x}_2=x_1-x_2$ & $\dot{x}_3=x_1+0.3x_3$\\
       ${\rm L}$ & $\dot{x}_1=x_2+3.9x_3$ & $\dot{x}_2=0.9x_1^2-x_2$ & $\dot{x}_3=1-x_1$\\
       ${\rm M}$ & $\dot{x}_1=-x_3$ & $\dot{x}_2=-x_1^2-x_2$ & $\dot{x}_3=1.7+1.7x_1+x_2$\\
       ${\rm N}$ & $\dot{x}_1=-2x_2$ & $\dot{x}_2=x_1+x_3^2$ & $\dot{x}_3=1+x_2-2x_1$\\
       ${\rm O}$ & $\dot{x}_1=x_2$ & $\dot{x}_2=x_1-x_3$ & $\dot{x}_3=x_1+x_1\star x_3+2.7x_2$\\
       ${\rm P}$ & $\dot{x}_1=2.7x_2+x_3$ & $\dot{x}_2=-x_1+x_2^2$ & $\dot{x}_3=x_1+x_2$\\
       ${\rm Q}$ & $\dot{x}_1=-x_3$ & $\dot{x}_2=x_1-x_2$ & $\dot{x}_3=3.1x_1+x_2^2+0.5x_3$\\
       ${\rm R}$ & $\dot{x}_1=0.9-x_2$ & $\dot{x}_2=0.4+x_3$ & $\dot{x}_3=x_1\star x_2-x_3$\\
       ${\rm S}$ & $\dot{x}_1=-x_1-4x_2$ & $\dot{x}_2=x_1+x_3^2$ & $\dot{x}_3=1+x_1$\\
\hline
\end{tabular}

\begin{center}
{\bf Table 3.} The noncommutative Sprott systems.
\end{center}
~~\\
We see, that due to the formulas (\ref{eq3}) and (\ref{eq4}), the above models (apart of Sprott-F, H, I, J, L, M, N, P, Q and S systems)
become complex; let us selectively examine them. In this aim we take under considerations the most simple (canonical) type of space noncommutativity (\ref{eq1}) given by $f_{ij}(t) = \theta_{ij} = {\rm const.}$ Then, there appears in all modified Sprott models the additional constant and pure imaginary term $\frac{i}{2}{\theta_{ij}}$. Below, we derive exemplary 3D and in $x_1$-direction trajectories as well as Poincar\'{e} sections for deformed Sprott-A system with parameter $\theta_{23}=1$, and with real initial conditions: $(x_1(0),x_2(0),x_3(0))=(0,5,0)$, $(x_1(0),x_2(0),x_3(0))=(0,5+\epsilon_1,0)$ and $(x_1(0),x_2(0),x_3(0))=(0,5+\epsilon_2,0)$ for $\epsilon_1 = 0.00001$ and $\epsilon_2 = 0.000001$ respectively\footnote{Trajectories are given by real component of the complex solution of differential Sprott equations, i.e., it looks as follows: $({\rm Re}(x_1(t)),{\rm Re}(x_2(t)),{\rm Re}(x_3(t)))$.}. The obtained results are presented on {\bf Figures 1 - 7} and they indicate, that the model remains sensitive with respect already small change of initial data. Besides, the shape of all three-dimensional trajectories as well as the form of the Poincar\'{e} sections suggest, that the investigated system irregularly oscillates along the stretch of the real $x_1$-axis\footnote{The numerical studies have been performed on the ${\it Mathematica}$ program version 10.1, with the accuracy and precision both equal to 16.}.

Further, we find by tedious calculations the
corresponding (noncommutative) jerk dynamics listed in the following two Tables:
\\
\\
\\
\begin{tabular}{ccccc}
\hline $\;\;{\rm type}\;\;$ & $\;\;\;\;\;\;\;\;\;\;\;\;\;\;\;\;\;\;\;\;\;\;\;\;\;\;\;\;\;\;\;\;\;\;\;\;\;\;\;\;\;\;\;\;\;
{\rm jerk\;dynamics}\;\;\;\;\;\;\;\;\;\;\;\;\;\;\;\;\;\;\;\;\;\;\;\;\;\;\;\;\;\;\;\;\;\;\;\;\;\;\;\;\;\;\;\;\;$ &  $$\\
\hline ${\rm A}$ & $-----------$ \\
       ${\rm B}$ & $-----------$ \\
       ${\rm C}$ & $-----------$ \\
       ${\rm D}$ & $\dddot{x}_1=x_1\ddot{x}_1-\dot{x}_1-3\dot{x}_1^2+x_1^2 - \frac{if_{13}(t)}{2}$ \\
       ${\rm E}$ & $-----------$ \\
       ${\rm F}$ & $\dddot{x}_1=-0.5\ddot{x}_1-0.5\dot{x}_1-0.5x^2_1+2\dot{x}_1x_1-x_1$\\
       ${\rm G}$ & $\dddot{x}_1=-0.6\ddot{x}_1-0.6\dot{x}_1-0.4x^2_1+\dot{x}_1x_1-x_1+\frac{if_{13}(t)}{2}$\\
       ${\rm H}$ & $\dddot{x}_3=-0.5\ddot{x}_3-0.5\dot{x}_3-0.5x^2_3+2\dot{x}_3x_3-x_3$\\
       ${\rm I}$ & $\dddot{x}_1=-\ddot{x}_1-0.2\dot{x}_1-5\dot{x}_1^2-0.4x_1$\\
       ${\rm J}$ & $\dddot{x}_2=-2\ddot{x}_2-\dot{x}_2-4x_2+2\dot{x}_2x_2$\\
       ${\rm K}$ & $\dddot{x}_2=-0.7\ddot{x}_2-0.7\dot{x}_2-x_2+x_2\ddot{x}_2+1.7x_2\dot{x}_2-0.3x_2^2+\dot{x}_2^2\;+$\\
       $~~$ & $\;\;\;\;\;\;\;\;\;\;\;\;\;\;\;\;\;\;\;\;\;\;\;\;\;\;\;\;+\;\frac{i}{2}\left(\dot{f}_{12}(t)-0.3{f_{12}(t)}\right)$\\
       ${\rm L}$ & $\dddot{x}_1=-\ddot{x}_1-3.9\dot{x}_1-3.9x_1+1.8x_1\dot{x}_1+3.9$\\
       ${\rm M}$ & $\dddot{x}_1=-\ddot{x}_1-1.7\dot{x}_1-1.7x_1+x_1^2-1.7$\\
       ${\rm N}$ & $\dddot{x}_3=-2\ddot{x}_3-2\dot{x}_3-4x_3+2\dot{x}_3x_3+2$\\
       ${\rm O}$ & $\dddot{x}_1=x_1\ddot{x}_1-1.7\dot{x}_1-x_1-x_1^2-\frac{if_{13}(t)}{2}$\\
       ${\rm P}$ & $\dddot{x}_2=2x_2\ddot{x}_2-1.7\dot{x}_2+2\dot{x}_2^2-x_2-x_2^2$\\
       ${\rm Q}$ & $\dddot{x}_2=-0.5\ddot{x}_2-2.6\dot{x}_2-3.1x_2-x_2^2$\\
       ${\rm R}$ & $\dddot{x}_1=-\ddot{x}_1-0.9x_1 +x_1\dot{x}_1-0.4-\frac{if_{12}(t)}{2}$\\
       ${\rm S}$ & $\dddot{x}_3=-\ddot{x}_3-4\dot{x}_3-4x_3^2+4$\\
\hline
\end{tabular}

\begin{center}
{\bf Table 4a.} The noncommutative jerk dynamics.
\end{center}
~~\\
\begin{tabular}{ccccc}
\hline $\;\;{\rm type}\;\;$ & $\;\;\;\;\;\;\;\;\;\;\;\;\;\;\;\;\;\;\;\;\;\;\;\;\;\;\;\;\;\;\;\;\;\;\;\;\;\;\;
{\rm remaining\;trajectories}\;\;\;\;\;\;\;\;\;\;\;\;\;\;\;\;\;\;\;\;\;\;\;\;\;\;\;\;\;\;\;\;\;\;\;\;\;\;\;$ &  $$\\
\hline ${\rm A}$ & $-----------$ \\
       ${\rm B}$ & $-----------$ \\
       ${\rm C}$ & $-----------$ \\
       ${\rm D}$ & $x_2=-\dot{x}_1\;\;\;\;\;\;\;\;\;\;\;\;\;\;\;\;\;\;\;\;\;\;\;\;\;\;\;\;\;\;\;\;\;x_3=-\ddot{x}_1-x_1$ \\
       ${\rm E}$ & $-----------$ \\
       ${\rm F}$ & ${x}_2=0.6(\ddot{x}_1+\dot{x}_1+x_1-{x}_1^2)\;\;\;\;\;\;\;\;\;\;{x}_3=0.6(-\ddot{x}_1+0.5\dot{x}_1-x_1+{x}_1^2)$\\
       ${\rm G}$ & ${x}_2=\ddot{x}_1-0.4\dot{x}_1+x_1\;\;\;\;\;\;\;\;\;\;\;\;\;\;\;\;\;\;\;\;\;\;\;\;\;\;x_3=\dot{x}_1-0.4x_1\;\;\;\;\;\;\;\;\;$\\
       ${\rm H}$ & ${x}_1=\dot{x}_3+{x}_3\;\;\;\;\;\;\;\;\;\;\;\;\;\;\;\;\;\;\;\;\;\;\;\;\;\;\;\;\;x_2=-\ddot{x}_3-\dot{x}_3+x^2_3$\\
       ${\rm I}$ & $\;\;\;\;\;{x}_2=-5\dot{x}_1\;\;\;\;\;\;\;\;\;\;\;\;\;\;\;\;\;\;\;\;\;\;\;\;\;\;\;\;\;\;\;\;\;\;\;\;x_3=-5\ddot{x}_1-x_1$\;\;\;\;\;\;\;\;\;\\
       ${\rm J}$ & $\;\;\;{x}_1=-\ddot{x}_2-2\dot{x}_2+x_2+{x}_2^2\;\;\;\;\;\;\;\;\;\;\;\;\;\;\;\;\;\;\;\;\;\;\;\;\;\;\;x_3=2x_2+\dot{x}_2\;\;\;\;\;\;\;\;\;\;\;\;\;\;\;\;$\\
       ${\rm K}$ & $\;\;\;\;\;{x}_1=\dot{x}_2+{x}_2\;\;\;\;\;\;\;\;\;\;\;\;\;\;\;\;\;\;\;\;\;\;\;\;\;\;\;\;\;\;\;x_3=-\ddot{x}_2-\dot{x}_2+x_2\dot{x}_2+x_2^2+
       \frac{if_{12}(t)}{2}$\\
       ${\rm L}$ & ${x}_2=-\ddot{x}_1-3.9{x}_1+0.9x_1^2+3.9\;\;\;\;\;\;\;\;\;\;\;\;\;\;\;x_3=0.2(\ddot{x}_1+\dot{x}_1+3.9x_1\;+$\\
       $~~$ & $\;\;\;\;\;\;\;\;\;\;\;\;\;\;\;\;\;\;\;\;\;\;\;\;\;\;\;\;\;\;\;\;\;\;\;\;\;\;\;\;\;\;\;\;\;\;\;\;\;\;\;\;\;\;\;\;\;\;-\;0.9x_1^2-3.9)$\\
       ${\rm M}$ & ${x}_2=-\ddot{x}_1-1.7x_1-1.7\;\;\;\;\;\;\;\;\;\;\;\;\;\;\;\;\;\;\;\;\;\;\;\;x_3=-\dot{x}_1\;\;\;\;\;\;\;\;\;\;\;\;\;\;$\\
       ${\rm N}$ & ${x}_1=\ddot{x}_3+2\dot{x}_3-{x}_3^2\;\;\;\;\;\;\;\;\;\;\;\;\;\;\;\;\;\;\;\;\;\;\;\;\;\;\;\;\;\;x_2=\dot{x}_3+2x_3-1$\\
       \end{tabular}
\eject
$~~$\\
\\
\begin{tabular}{ccccc}
$\;\;\;\;\;\;\;\;\;\;\;\;\;\;$ &
$\;\;\;\;\;\;\;\;\;\;\;\;\;\;\;\;\;\;\;\;\;\;\;\;\;\;\;\;\;\;\;\;\;\;\;\;\;\;\;\;\;\;\;\;\;\;\;\;\;\;\;\;\;\;\;\;\;\;\;\;
\;\;\;\;\;\;\;\;\;\;\;\;\;\;\;\;\;\;\;\;\;\;\;\;\;\;\;\;\;\;\;\;\;\;\;\;\;\;\;\;\;\;\;\;\;\;\;\;\;$ &  $$\\
       ${\rm O}$ & $\;\;\;\;\;{x}_2= \dot{x}_1\;\;\;\;\;\;\;\;\;\;\;\;\;\;\;\;\;\;\;\;\;\;\;\;\;\;\;\;\;\;\;\;\;\;\;\;x_3=-\ddot{x}_1+x_1$\;\;\;\;\;\;\;\\
       ${\rm P}$ & $\;\;\;\;\;\;{x}_1=-\dot{x}_2+x_2^2\;\;\;\;\;\;\;\;\;\;\;\;\;\;\;\;\;\;\;\;\;\;\;\;\;\;\;\;\;\;\;\;\;x_3=-\ddot{x}_2+2x_2\dot{x}_2-2.7x_2$\\
       ${\rm Q}$ & $\;\;\;\;\;{x}_1=\dot{x}_2+{x}_2\;\;\;\;\;\;\;\;\;\;\;\;\;\;\;\;\;\;\;\;\;\;\;\;\;\;\;\;\;\;\;\;\;\;\;\;x_3=-\ddot{x}_2-\dot{x}_2$\;\;\;\;\;\;\;\;\;\\
       ${\rm R}$ & $\;\;\;\;\;{x}_2=-\dot{x}_1+0.9\;\;\;\;\;\;\;\;\;\;\;\;\;\;\;\;\;\;\;\;\;\;\;\;\;\;\;\;\;\;\;\;\;\;\;\;x_3=-\ddot{x}_1-0.4$\;\;\;\;\;\;\;\;\;\\
       ${\rm S}$ & $\;\;\;\;\;\;\;\;\;\;{x}_1=\dot{x}_3-1\;\;\;\;\;\;\;\;\;\;\;\;\;\;\;\;\;\;\;\;\;\;\;\;\;\;\;\;\;\;\;\;\;x_2=0.25(-\ddot{x}_3-\dot{x}_3+1)$\\
\hline
\end{tabular}

\begin{center}
{\bf Table 4b.} The noncommutative jerk dynamics - remaining trajectories.
\end{center}
Consequently, in such a way, we easily observe that all founded above complex equations include additional pure imaginary terms $\frac{i}{2}f_{ij}(t)$ and $\frac{i}{2}\dot{f}_{ij}(t)$ respectively, and just for this reason, in the case of functions $f_{ij}(t)$ equal zero, the discussed models become classical.

At the end stage of this paper, it should be also mentioned, that the physical meaning of all complex components Im$(x_i(t))$ remains unclear. However, one may expect for example, that they play a role of additionally, generated by space-time noncommutativity, internal degrees of freedom. They are not observed or directly measured, but in spite of that, they effectively affect on the real part of the solution of differential equation.

\eject
\pagestyle{empty}
$~~~~~~~~~~~~~~~~~~$
\\
\\
\\
\\
\\
\\
\\
\\
\\
\begin{figure}[htp]
\includegraphics[width=\textwidth]{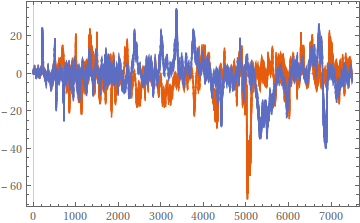}
\caption{The Re$(x_1(t))$-trajectories for Sprott-A system with initial conditions $(x_1(0),x_2(0),x_3(0)) = (0,5,0)$ (red line) and $(x_1(0),x_2(0),x_3(0)) = (0,5+\epsilon_1,0)$ (blue line) respectively. The time parameter runs from 0 to 7500 and the trajectories become divergent near $t$ equal to 500.}\label{grysunek1}
\end{figure}
\eject
\pagestyle{empty}
$~~~~~~~~~~~~~~~~~~$
\\
\\
\\
\\
\\
\\
\\
\\
\\
\begin{figure}[htp]
\includegraphics[width=\textwidth]{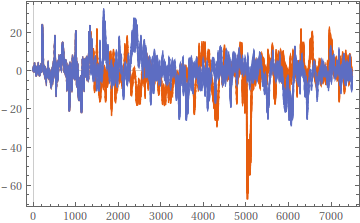}
\caption{The Re$(x_1(t))$-trajectories for Sprott-A system with initial conditions $(x_1(0),x_2(0),x_3(0)) = (0,5,0)$ (red line) and $(x_1(0),x_2(0),x_3(0)) = (0,5+\epsilon_2,0)$ (blue line) respectively. The time parameter runs from 0 to 7500 and the trajectories become divergent near $t$ equal to 1200.}\label{grysunek1a}
\end{figure}
\eject
\pagestyle{empty}
$~~~~~~~~~~~~~~~~~~$
\\
\\
\\
\\
\\
\\
\\
\\
\\
\begin{figure}[htp]
\includegraphics[width=\textwidth]{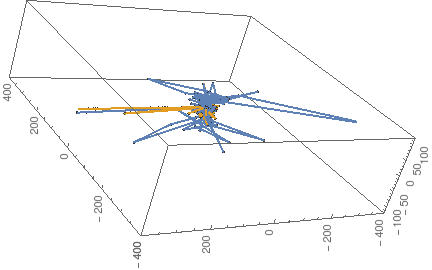}
\caption{The 3D-trajectories for Sprott-A system with initial conditions $(x_1(0),x_2(0),x_3(0)) = (0,5,0)$ (blue line) and $(x_1(0),x_2(0),x_3(0)) = (0,5+\epsilon_1,0)$ (orange line) respectively. The time parameter runs from 0 to 20000.}
\label{grysunek3}
\end{figure}
\eject
\pagestyle{empty}
$~~~~~~~~~~~~~~~~~~$
\\
\\
\\
\\
\\
\\
\\
\\
\\
\begin{figure}[htp]
\includegraphics[width=\textwidth]{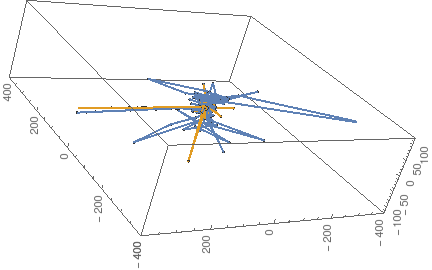}
\caption{The 3D-trajectories for Sprott-A system with initial conditions $(x_1(0),x_2(0),x_3(0)) = (0,5,0)$ (blue line) and $(x_1(0),x_2(0),x_3(0)) = (0,5+\epsilon_2,0)$ (orange line) respectively. The time parameter runs from 0 to 20000.}\label{grysunek4a}
\end{figure}
\begin{figure}[htp]
\includegraphics[width=\textwidth]{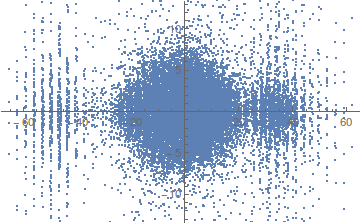}
\caption{The Poincar\'{e} section (Re$(x_1(t))$,Re$(x_3(t))$) with Re$(x_2(t))=0$ for Sprott-A model for initial condition $(x_1(0),x_2(0),x_3(0)) = (0,5,0)$. The time parameter runs from 0 to 20000.}\label{grysunek3a}
\end{figure}
\begin{figure}[htp]
\includegraphics[width=\textwidth]{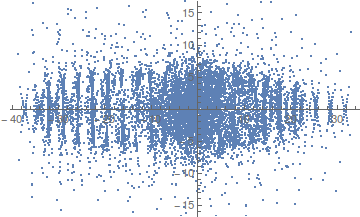}
\caption{The Poincar\'{e} section (Re$(x_1(t))$,Re$(x_3(t))$) with Re$(x_2(t))=0$ for Sprott-A system for initial condition $(x_1(0),x_2(0),x_3(0)) = (0,5+\epsilon_1,0)$. The time parameter runs from 0 to 20000.}\label{grysunek4}
\end{figure}

\begin{figure}[htp]
\includegraphics[width=\textwidth]{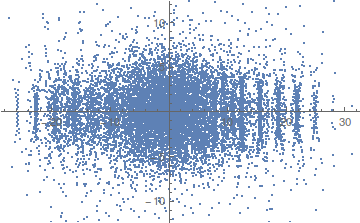}
\caption{The Poincar\'{e} section (Re$(x_1(t))$,Re$(x_3(t))$) with Re$(x_2(t))=0$ for Sprott-A system for initial condition $(x_1(0),x_2(0),x_3(0)) = (0,5+\epsilon_2,0)$. The time parameter runs from 0 to 20000.}\label{grysunek4}
\end{figure}

\end{document}